# Individuality and slow dynamics in bacterial growth homeostasis

Lee Susman[1,2], Maryam Kohram[3], Harsh Vashistha[3], Jeffrey T. Nechleba[3], Hanna Salman[3,4,*], and Naama Brenner[1,5,*]

[1]Network Biology Research Laboratory, Technion-Israel Institute of Technology, Haifa 32000, Israel
[2]Interdisciplinary Program of Applied Math, Technion-Israel Institute of Technology, Haifa 32000, Israel
[3]Department of Physics and Astronomy, University of Pittsburgh, Pittsburgh PA 15260, USA
[4]Department of Computational and Systems Biology, University of Pittsburgh, Pittsburgh PA 15260, USA
[5]Department of Chemical Engineering, Technion-Israel Institute of Technology, Haifa 32000, Israel
*Corresponding Authors: nbrenner@technion.ac.il, hsalman@pitt.edu

**Abstract**

Microbial growth and division are fundamental processes relevant to many areas of life science. Of particular interest are homeostasis mechanisms, which buffer growth and division from accumulating fluctuations over multiple cycles. These mechanisms operate within single cells, possibly extending over several division cycles. However, all experimental studies to date have relied on measurements pooled from many distinct cells. Here, we disentangle long-term measured traces of individual cells from one another, revealing subtle differences between temporal and pooled statistics. By analyzing correlations along up to hundreds of generations, we find that the parameter describing effective cell-size homeostasis strength varies significantly among cells. At the same time, we find an invariant cell size which acts as an attractor to all individual traces, albeit with different effective attractive forces. Despite the common attractor, each cell maintains a distinct average size over its finite lifetime with suppressed temporal fluctuations around it, and equilibration to the global average size is surprisingly slow ($> 150$ cell cycles). To demonstrate a possible source of variable homeostasis strength, we construct a mathematical model relying on intracellular interactions, which integrates measured properties of cell size with those of highly expressed proteins. Effective homeostasis strength is then influenced by interactions and by noise levels, and generally varies among cells. A predictable and measurable consequence of variable homeostasis strength appears as distinct oscillatory patterns in cell size and protein content over many generations. We discuss the implications of our results to understanding mechanisms controlling division in single cells and their characteristic timescales.

## Significance statement

Microbial cells go through repeated cycles of growth and division. These cycles are not perfect: the time and size at division can fluctuate from one cycle to the next. Still, cell size is kept tightly controlled and fluctuations do not accumulate to large deviations. How this control is implemented in single cells is still not fully understood.

We performed experiments that follow individual bacteria in microfluidic traps for hundreds of generations. This enables us, for the first time, to identify distinct individual dynamic properties that are maintained over many cycles of growth and division. Surprisingly, we find that each cell suppresses fluctuations with a different strength; this variability defines an "individual" behavior for each cell, which is inherited along many generations.

\body

## Introduction

The processes of growth and division in proliferating cells have been of interest for decades, with microorganisms providing model systems for both experimental and theoretical studies. Recently, with the development of experimental methods (1), new light was shed on this problem. Large samples of cells can be continuously tracked as they grow and divide for multiple cycles at high spatial and temporal resolution. Such measurements provide ample new information about these processes. As an important quantitative result of single-cell tracking, it was shown that individual yeast cells (2), as well as different bacterial cells (3–6), grow exponentially in time to a good approximation between consecutive divisions. This result sharpens the problem of cell-size homeostasis, since successive cycles of exponential growth and division can be unstable to fluctuations (7–11). Negative correlations can, in principle, prevent the instability and divergence caused by independent fluctuations. Indeed, using large samples of growth and division cycles pooled from single cell measurements, a negative correlation was found between the size change over the cell-cycle and the initial size (i.e. cell size at the start of the cell cycle) (7, 12–17). The observed correlations, if interpreted as regulation of cell division, rule out two previously studied models of division control, namely, a constant fold-change over the cell cycle, and a fixed size-threshold for division. However, despite these new data and insights, a clear mechanism linking cell size to division has not yet been identified (18).

From a theoretical modelling perspective, the experimentally observed correlations provide the basis for a phenomenological approach, without reference to any specific underlying mechanism (7, 8, 17–19). Such an approach can be formulated mathematically in several essentially equivalent ways (20), all of which incorporate the exponential accumulation and its above-mentioned negative correlation with the initial cell size. Although this correlation does not necessarily imply a causal relation, it is convenient to envision it as a restraining force which counteracts the fluctuating exponentials, attracting them to the bottom of an effective potential well (20, 22). In this picture, a cell born too large will accumulate a smaller fold-change over the growth cycle and vice versa, preventing fluctuations from accumulating in the long run.

Several experiments in *E. coli* and *S. cerevisiae* were found to be consistent with a specific value of the effective force constant. This particular value corresponds, in a linear approximation, to a fixed volume added on average at each cycle, and was hence termed the "adder" model (8, 11, 12, 19, 21). Closer inspection of the data, however, reveals that the correlation plots are very noisy, despite the large samples and high accuracy of the experiments. Moreover, some experiments showed force constants different from the one corresponding to the adder model. Investigation of *E. coli* and mycobacteria in different environments, for example, resulted in a range of different measured values (15, 16). Experiments in the *C. crescentus* revealed two phases in the cell cycle, each characterized by a different restraining force strength (17).

All these previous studies have used measurements pooled from many single cells to increase statistics. Such pooled data can provide information on cell-cycle parameters averaged over the entire ensemble of cells; however, mechanisms that regulate and control division operate at the level of the single cell, and their individual properties might be masked by such pooling. In the current study, we measure and analyze dynamics of growth and division in individual bacteria tracked over extended times, up to ~250 cell cycles each. Making the distinction between statistics over time in individual cells and the corresponding statistics averaged over many cells requires, first, that long enough stable individual traces be acquired without confounding effects, such as filamentation or contamination; and second, that statistical properties be analyzed from separate traces and compared to those averaged over many traces. Our previous work carried out such a comparison for protein distributions and found a universality of distribution shape in both ensembles (10). To make a similar comparison for cell-size homeostasis, which is a dynamic process, longer traces and more statistics are required. Here we present data that enables this comparison.

Our results show that individual cells exhibit different values of the effective restraining force constant, which is maintained distinct for many cell cycles. At the same time, an invariant is revealed in the form of an ensemble-average cell-size, acting as an attractor to the dynamics over long times. Despite this common attractor, we find significant differences in temporally averaged size between traces over the finite lifetime of each cell. This is related to deviations of temporally-averaged division ratio and fold-change from their global average values of 1/2 and 2, respectively. Such deviations are persistent over dozens of division cycles, and equilibration to the global averages appears only in the longest traces, those over 150 cycles long.

Integrating cell-size data with measurements of protein content in the same cells, we propose that a possible origin of variable homeostasis strength stems from underlying interactions between global cellular variables. We present an illustrative mathematical model of these interactions, which reproduces several non-trivial aspects of the entire dataset. As a consequence of the individuality in homeostasis parameter (restraining force constant), we provide a theoretical explanation for oscillatory autocorrelations in cell size and in protein content, which have been previously reported (15). We discuss the implications of our results to the quest for the mechanisms underlying cellular growth and division homeostasis, and point to new future research directions.

## Results

### Cell-size homeostasis: single cell vs. ensemble-average behavior

Continuous measurements of cell-size over time reveal smooth exponential-like accumulation throughout each cell cycle, interrupted by abrupt drops at division. Fig. 1A shows a small portion of such a measurement. Cell length is taken as an attribute of cell size, as the rod-like bacteria grow in one dimension along their length, while their width is maintained constant (see Methods; (3, 10)). Over the $n$-th cycle of growth and division, cell size $x_n(t)$ can be described accurately as:

$$x_n(t) = x_n(0) e^{\alpha_n t}, \quad 0 < t < T_n$$
$$x_{n+1}(0) = f_n x_n(T_n),$$

where $\alpha_n$ is the exponential accumulation rate during cell-cycle $n$; $T_n$ its duration; and $f_n$ the division fraction at its end (see black fitting line in Fig. 1A). The cell size at the start of the $n$-th cell cycle $x_n(0)$, which we denote simply as $x_n$, is linked across generations by the mapping:

$$x_{n+1} = f_n x_n e^{\phi_n}, \tag{1}$$

with the total accumulation exponent defined as $\phi_n = \alpha_n T_n$ (7). In this discrete mapping, each step represents a complete cell cycle and is characterized by two variables: a total increase in cell size by a factor $e^{\phi_n}$ from beginning to end of the cycle, and a decrease by a factor of $f_n$ at division. The definition of these parameters is illustrated in Fig. 1A. Both increase and decrease are subject to fluctuations from one cell cycle to the next (Fig. SI-1-1 shows their distributions); cell size homeostasis requires their product to average to one over long times. Pooling together a large sample of cell cycles from many individual traces shows that, on average, this is indeed the case (Fig. SI-1-1). However, while this requirement is necessary, it is insufficient for homeostasis: the process described by Eq. (1) is unstable against fluctuations in $f_n$ and $e^{\phi_n}$ over long times, even if on average their product is 1; independent fluctuations accumulate and the variance increases with time. Fig. 1B shows that the exponential accumulation of size during a cell-cycle, $\phi_n$, is negatively correlated with initial cell size. Such state-dependent changes can control fluctuations and induce a stable size distribution over multiple generations (7,8; see SI-2). Additional empirical correlations are presented in Fig. SI-1-2.

To demonstrate the effect of state-dependent changes on long-term dynamics, we present in Fig. 1C a comparison between two sequences of cell sizes $x_n$: the first (blue) is an actual measurement of a single cell for over 200 generations, while the second (gray) is created artificially by randomly shuffling the order of the pairs $(\phi_n, f_n)$ in the same trace. Such a permutation retains the average of the products $e^{\phi_n} f_n$ at 1, but loses the correlations displayed in Fig. 1B. Clearly, these correlations contribute to the stabilization of cell size, limiting it to a small range.

One may account for the negative correlations of Fig. 1B in the mapping (Eq. (1)) by postulating a relation between $x_n$ and the fold-change $e^{\phi_n}$ (7, 8). Consistent with the data, we may use a linear approximation in logarithmic coordinates:

$$\phi_n = \phi^* - \beta \ln \frac{x_n}{x^*} + \xi_n, \qquad (2)$$

where $\beta$ represents the slope of the fit in Fig. 1B. We term this the homeostasis parameter. The noise, $\xi_n$, has zero mean and is approximately Gaussian. Here the typical size $x^*$ sets the units in which cell size is measured. This scaling size $x^*$ is chosen to be the average cell size at the start of the cell-cycle over the entire data-set, or the 'ensemble average' cell size. We find that the best linear fit is obtained with $\phi^* \approx \ln 2$, corresponding to a mean fold-change of 2, as expected, when averaging over many cells with a mean division ratio of ½ (see Fig. SI-1-1).

We next consider the same analysis of correlations between cell-cycle variables, applied to individual traces separately; a similar picture may be expected, but with some degree of variability among individual cells. This variability could be due to noise, in which case they will have similar linear correlation parameters, up to errors resulting from measurement noise and finite sampling. On the other hand, significant variability in the correlations could reflect true individuality of cells. We write the analog of Eq. (2) for the $k$-th individual trace,

$$\phi_n^{(k)} = \phi^{*(k)} - \beta^{(k)} \ln \frac{x_n^{(k)}}{x^*} + \xi_n^{(k)}, \qquad (3)$$

with cell size still measured in units of the ensemble average $x^*$. Variability can be reflected as significant differences in any one of the parameters $\phi^{*(k)}, \beta^{(k)}$, or the properties of the noise $\xi_n^{(k)}$.

In graphical terms, the pooled scatterplot of Fig. 1B could be composed of single-cell plots that differ in their properties in several alternative ways; however, they could also exhibit invariant features common to all traces (see Box 1 for illustration). Identifying those properties that are conserved among all cells may point to their importance as control variables.

Fig. 2A highlights two individual measured traces in color (green and red) on the background of the entire ensemble in gray, suggesting that they have distinct values of the homeostasis parameter. Fig. 2B shows the estimated slopes $\beta^{(k)}$ for all traces as a function of trace length in number of generations. The error-bars, representing the uncertainty in the slope, are significantly smaller than the differences between traces for traces of length $< 150$ generations; these differences decrease for the extremely long traces of length $150 - 250$ generations. To quantify the difference between traces, two methods of statistical analysis were applied, showing that the apparent difference is statistically significant beyond the noise and the finite sampling (see SI-3).

Fig. 2C shows the best linear fits for all individual traces. Each black line was obtained as a fit similar to the colored lines in Fig. 2A, taken from a single individual trace along time. This figure reveals an invariant in the form of a pivot point where all lines cross. The coordinates of this point coincide with the ensemble average (green circle in Fig. 2C), implying a common intercept $\phi^{*(k)} = \phi^* = \ln 2$ for all traces. This leaves one parameter, the homeostatic parameter or correlation slope $\beta^{(k)}$, which is distinct to each trace.

The pivot point, common to all individual traces, suggests a dynamic attractor for cell-size over multiple cycles. In this picture, if the cell divides to a size that considerably deviates from $x^*$, the exponential accumulation during the following cell cycle would be compensated to effectively "pull" the cell back to this common attractor, with a force strength variable among cells. This picture is supported by Fig. 2D, in which a flow map computed as an average over all cycles in the data-set is presented. Similar dynamics are found also in other experimental conditions, for example a different nutrient composition (SI-4). We note that, while $\phi^* = \ln 2$ for all conditions tested, the ensemble-average cell size at the start of the cell-cycle, $x^*$, depends on growth medium and temperature (see SI-4).

**Differences between time-averaged cell sizes of individual traces**

Do the distinct values of $\beta^{(k)}$ result in different cell-sizes, when averaged over the lifetime of the cell? We use the mapping model for individual traces to answer this question: combining the two equations (1) and (3), the mapping can be written as

$$\ln \frac{x_{n+1}^{(k)}}{x^*} = \left(1 - \beta^{(k)}\right) \ln \frac{x_n^{(k)}}{x^*} + \ln f_n^{(k)} + \phi^* + \xi_n^{(k)}, \tag{4}$$

linking the logarithm of cell size in consecutive cell cycle starts. Here we have incorporated the empirical observations that homeostasis parameters are distinct, whereas the intercept $\phi^*$ is common to all traces. Denoting temporal averaging over a trace by overbars, we compute from Eq. (4) the time averaged logarithm of initial cell-size:

$$\overline{\ln x_n^{(k)}} = \ln x^* + \frac{\phi^* + \overline{\ln f_n^{(k)}}}{\beta^{(k)}}, \tag{5}$$

where we have used the empirical result $\overline{\xi_n^k} \approx 0$. If $\overline{\ln f_n^{(k)}} = \ln \frac{1}{2}$, the right term vanishes: $\phi^* + \overline{\ln f_n^{(k)}} = \ln 2 + \ln \frac{1}{2} = 0$; then, distinct values of $\beta^{(k)}$ may affect the rate of relaxation toward the attractor, but not the steady state itself. However, when we examine experimentally measured sequences of consecutive initial cell sizes $x_n$, we find that their averages are distinct. Fig. 3A displays such sequences for two long and stable traces, with horizontal lines depicting their temporal averages. The distribution of values along the trace is plotted on the right of Fig. 3A, for each trace with its corresponding color.

To understand the origin of these differences, we simulated two traces using the model, Eq. (4), with homeostasis parameters $\beta^{(k)}$ ($k = 1, 2$) taken from the two traces in Fig. 3A, and with $\phi^* = \ln 2$. Both $\xi_n$ and $f_n$ were simulated as random variables drawn independently at each step, with statistical properties matching those of the entire data-set, i.e. $\langle \xi_n \rangle = 0$, and $\langle \ln f_n \rangle = \ln 1/2$. Fig. 3B shows the two simulated traces. In comparison, the measured traces in Fig. 3A exhibit suppressed temporal fluctuations, each around a different mean value, farther removed from one another than the model predicts. These effects can be quantified by computing the standard deviation "internal" to an individual trace, $\sigma_{\text{in}}(\ln(x/x^*))$, estimated over time (width of distributions on the right, Figs. 3A, 3B; see Methods). We find that, on average over all measured traces, $\langle \sigma_{\text{in}}(\ln(x/x^*)) \rangle = 0.26 \pm 0.06$, while for a corresponding collection of simulated traces, we find $0.43 \pm 0.05$. On the other hand, fluctuations of the measured traces are centered around temporal averages which are significantly different from one another. This can be quantified by the "external" standard deviation of time-averaged sizes, $\overline{\ln(x/x^*)}$, across all traces: we find that $\langle \sigma_{\text{ex}}(\overline{\ln(x/x^*)}) \rangle = 0.14$ for experimental data and $0.06$ for the model simulation results (see Methods for details). This analysis provides statistical support to the effect apparent in Fig. 3A, namely that cell size fluctuations along time are strongly suppressed in each trace around a distinct time-averaged value.

The discrepancy between individual traces and the model prediction suggest a distinct behavior of the division ratio in each trace: in the model, this was taken as a random variable common to all traces and drawn from a Gaussian distribution around 1/2. However, each cell undergoes a limited number of growth and division cycles before it dies. Therefore, if division ratios maintain a bias that deviates on average from 1/2 along many cycles, the effective feedback in the exponential accumulation $\phi_n$ – which ensures balanced growth – induces a corresponding deviation of the average fold-increase from 2. Consequently, the range of values sampled in the $(\ln x_n, \phi_n)$-plane by an individual trace over its finite lifetime may be biased and not provide a good sample of the range attained by the entire ensemble.

This signature of slow dynamics manifests as a distinct clustering of the points making up each trace, as illustrated in Fig. 3C. The clusters corresponding to the two individual traces, indeed, are not overlapping and each trace samples a slightly different portion of the space. Large colored circles depict the average of the colored points corresponding to two individual traces and illustrate how the distinct clusters result in distinct averages. The time-averaged division ratios and the corresponding time-averaged exponential accumulations are presented for all traces in Fig. 3D, demonstrating the extent of biases spanned by individual traces and the tight compensation between them induced by homeostasis. This explanation of distinct time-averaged sizes in terms of division-ratio bias agrees reasonably well with the data, as detailed in SI-5.

**Possible origin of variation in homeostasis parameters**

Why do individual cells exhibit distinct values of the homeostasis parameter, $\beta^{(k)}$? Recall that this parameter quantifies an empirical negative correlation between initial cell size and exponential size accumulation during the cell cycle. Mechanistically, little is known about underlying processes that may induce such a negative correlation. One might imagine that some molecular circuit implements a feedback loop from accumulated cell size to division; experiments have suggested different molecules to be implicated in such a process, but a specific mechanism has not yet been identified (18). Recently, several researchers have put forward the possibility that size homeostasis is not implemented at the molecular level, but may represent a global systems-level property of the cell ((18, 23, 24); see also (25)).

In line with this idea, we consider the dynamics of accumulation and division from a global cellular perspective. A key observation is that the copy number of highly expressed proteins, which also accumulates exponentially and divides over multiple cell cycles, exhibits an apparent 'protein homeostasis', reflected in negative correlations similar to cell size; namely, one may associate a nonzero value of $\beta$ with highly expressed proteins in the cell (7). These values, which are generally smaller than those corresponding to cell size, span a wide range for different proteins, conditions and individual cells (see SI-6).

Previous work has shown that traces of highly expressed protein content are not only qualitatively similar to those of cell size, but are statistically correlated with them on a cycle-by-cycle basis (12). To further characterize this relationship, we measured the copy-numbers of two fluorescent proteins simultaneously in single cells, along cycles of growth and division. One was expressed under the control of the *lac* promoter in a lactose rich medium and, thus, represents a metabolically relevant protein. The second was expressed under the control of the constitutive λ-phage PR promoter, which is foreign to the cell and does not contribute to cellular metabolism (see Methods for details). Examples of three simultaneously measured traces can be seen in Fig. 4A. A strong correlation between the three exponential rates measured in the same cell-cycle is seen in Fig. 4B.

Given these similarities and quantitative correlations, one may argue that the copy number of protein is simply proportional to cell-size. This would imply a constant (or narrowly distributed) protein density per unit volume. However, Fig. 4C shows that protein density spans a broad range, approximately 5-fold in concentration, suggesting that the relationship between protein and cell size is not a simple proportion. Furthermore, a model of several phenotype components which are 'enslaved' to cell size, accumulating with the same exponential rates and dividing at the same times controlled by cell size, is found to be unstable to fluctuations and cannot induce homeostasis on the entire multi-dimensional system (see SI-7-B).

If protein content is not enslaved to cell size and is not directly implicated in cell division control, why then does it have nonzero effective restraining force strength $\beta$? Taking a holistic view on cellular homeostasis, we consider the possibility that effective interactions between various measurable cellular characteristics (cell-size, protein content, etc.) give rise to an *effective* homeostasis parameter for protein content. We shall see that such a model also explains the variation among cells in the homeostasis parameter corresponding to cell size.

As a concrete implementation of this principle, we consider a set of $N$ cellular components, which we denote by a vector $\vec{x}$. To describe the effective interactions, we go beyond the mapping model, which only relates discrete time-points in consecutive generations, and include also dynamics of components within the cell cycle. Building on previous models of linear interactions, which give rise to indirectly auto-catalytic dynamics of all variables (26, 27), we write the equation of motion within cell-cycle $n$ as:

$$\frac{d}{dt}\vec{x_n}(t) = \mathcal{K}\vec{x_n}(t), \quad 0 < t < T_n, \tag{5}$$

where $\vec{x_n}(t)$ now describes the continuous evolution of all $N$ components. The effective interaction matrix $\mathcal{K}$ is randomly chosen and describes intracellular metabolism, fixed along cycles. Cell division distributes fractions $f_{j,n}$ and $(1 - f_{j,n})$ of component $j$ to each daughter cell,

$$x_{j,n+1}(0) = f_{j,n} \cdot x_{j,n}(T_n), \tag{6}$$

with $f_{j,n}$ randomly distributed around ½. The model description is completed by designating component 1 as controlling cell division, through the relation in Eq. (2) (see SI-7-C for other control strategies). Fig. 5A shows traces of three components in the same cell, resulting from numerically simulating the model. It shows that exponential-like accumulation and division persist stably over many generations in all components. The resulting picture is qualitatively insensitive to the number of components and to many properties of the interaction matrix $\mathcal{K}$ (see SI-7 for more details and conditions on the model).

The interactions within cell-cycles induce effective negative correlations with apparent homeostasis parameters for all components, thereby stabilizing long-term accumulation and division against fluctuations. As expected, phenotype component 1 exhibits a strong correlation between its accumulated exponent and its value at the start of the cell cycle (see Fig. 5B), since it actually implements the control of cell division. Perhaps more surprisingly, effective correlations $\tilde{\beta}_j^{(k)}$ emerge between accumulated exponents and initial values for all other components (Fig. 5C). This effective parameter varies among components $j$ of the simulation (here $(k)$ labels the individual trace as before): if interactions are strong enough (large off-diagonal matrix elements of $\mathcal{K}$), it can be as strong, or even stronger, than that of the controlling component. Furthermore, even for the controlling component itself, the interactions can modify the empirically measured effective correlation parameter such that it differs from the one originally assigned to it in division control ($\tilde{\beta}_1^{(k)} \neq \beta_1^{(k)}$). The effective $\tilde{\beta}_j^{(k)}$ are found to vary as a function of the interaction strengths specified by $\mathcal{K}$; for fixed interactions, they vary upon different realizations of noise at division (see Fig. SI-7-1C). When averaged over realizations, our model predicts its dependence on the noise properties. Cells with sloppy division (large $\sigma_f$, standard deviation of division fraction) but a sharp division condition (small $\sigma_\xi$, standard deviation of division control) have larger effective $\tilde{\beta}_1^{(k)}$, and vice versa (Fig. 5D).

To test these nontrivial model predictions, we return to the data and consider the dependence of the estimated $\tilde{\beta}_j^{(k)}$ of individual traces on the noise level, with $j = 1,2$ corresponding to cell-size and protein content. Our entire collection of traces spans a range of noise levels in their effective homeostatic correlation; in Fig. 5E we show the values of $\tilde{\beta}_j^{(k)}$ estimated for traces of both cell size (red) and protein content (green), as a function of $\sigma_\xi$. The dependence is qualitatively in agreement with the model predictions, depicted by the black line (horizontal projection of Fig. 5D). No systematic dependence was found as a function of noise in division $\sigma_f$, possibly due to the small variability in this parameter among individual traces.

The multi-component model with coupled dynamics reproduces many of the statistical properties of the experimental data described above, at the level of individual traces as well as the entire collection of traces (details in SI-7). Importantly, it provides a potential explanation for the emergence of a range of effective homeostasis parameters for different components, and for the variability among individual cells in the measured cell-size homeostasis parameter.

**Consequence of variable homeostasis parameter**

Recent work has shown that traces of cell size and protein content in single bacterial cells exhibit damped oscillations in their autocorrelation functions (ACFs) (15). These oscillations were linked through numerical simulations to a homeostatic mapping between consecutive generations, similar to the model used here. Fig. 6 displays the ACFs for several traces of cell size (A) and protein (B) showing these oscillations. Using the mapping model, we now show that the structure of these correlations is the predictable outcome of variable values of $\tilde{\beta}_j^{(k)}$ in individual cells and in different phenotype components (cell size, protein).

For any given $\beta$, computing the ACF via Eq. (4) by averaging over the ensemble from which noise is drawn, one finds an exponential with a time-constant of $\approx \ln(1-\beta)$ (in number of generations). This is in line with the smooth form that appears after averaging over all traces, where $\beta$ is some typical value in the ensemble (Figs. 6A, 6B black lines). In a single trace, on the other hand, one cannot directly calculate the ACF from the model. However, the probability of oscillatory patterns being generated at random across time and their period may be estimated (28; see SI-8). This probability depends on the trace-specific value of $\beta$, and therefore will be reflected in distinct oscillatory patterns.

Fig. 6C shows this theoretical prediction (black solid line) together with the corresponding quantities computed from our experimental traces (dots). Although the individual traces show a large scatter, binning them by value of $\tilde{\beta}_j^{(k)}$ agrees well with the theory (large circles). The scatter is expected, since the theory is probabilistic and predicts an average over realizations; it becomes a better predictor of model simulations as trace length increases (see SI-8). We see that the oscillatory patterns of the ACF arise from purely stochastic effects, in combination with the inherent discreteness of cell division, and that the individuality of the homeostasis parameter echoes in their distinct periods. This agreement of the theoretical prediction with the data provides an independent verification of the variability in $\tilde{\beta}_j^{(k)}$, specifically in the cell size homeostasis parameter, among individual traces.

## Discussion

The process of cellular growth and division is subject to many sources of noise, which can accumulate and lead to divergence over time if left unrestrained. In an effort to understand the restraining forces that maintain cell size homeostasis in bacteria, we have analyzed the size dynamics of many individual cells measured for up to hundreds of generations. Such dynamics can be described by a phenomenological model, with an effective feedback linking the exponential size accumulation during each cell-cycle, $\phi_n$, to the initial size in that cycle, $x_n$. This feedback, which is a negative correlation inferred directly from the data, acts as a restraining force for maintaining cell size from diverging over time. It has often been interpreted as a mode of division regulation, where specific restraining force strengths correspond to previously described regulation models – e.g. "adder", "sizer" etc. (see SI-2). Different modes display different slopes of the correlation in the $(\ln x_n, \phi_n)$ phase-space.

*Cell-size homeostasis: pooled cycles vs. individual traces.* When analyzing the data after pooling cycles from many individual traces, the correlation is consistent with a slope of ½, corresponding to the previously studied adder model. On the other hand, examining the data for each cell separately reveals that the correlation slopes vary from cell to cell. This variation is statistically significant beyond the noise in the measurements. Examining the entire collection of traces in our data-set, we find that the best linear fits cross at a common point, corresponding to the average cell size and the average exponential accumulation of ln2. The fact that all individual lines cross at one point is a nontrivial result; in principle, they could have varied in other ways that would remain consistent with the observed ensemble scatter-plot (see Box 1). This result points to a physiologically invariant cell size acting as a common attractor of the dynamics. Thus, in individual cells, homeostasis pulls against fluctuations towards a common cell size, albeit with different force strength. The actual value of the cell size at this attractor depends on experimental conditions.

Despite the common attractor, our measurements reveal that time-averaged cell sizes remain distinct among traces over dozens of generations. The difference between temporal averages of individual traces reflects slow dynamics that extend over this timescale. In particular, the exponential accumulation and division ratio do not always converge to the ensemble averages of 2 and ½, respectively, over the lifetime of the cell. This may seem as a surprising result; however, in principle, the existence of an effective feedback allows each trace to remain centered around distinct steady-state values without losing homeostasis. The mapping model can empirically predict these deviations reasonably well as stemming from slow dynamics of division ratios, whose temporal averages can deviate from ½ over many generations. Further work is required to understand how long-term deviations in a mother-cell are reconciled with the behavior of its daughter cells, and how eventually the lineages make up a population with symmetric division on average (29). These are topics for future research.

*Homeostasis of multiple cellular components.* To better understand homeostasis in individual cells, we examined not only the dynamics of their size but also of highly expressed proteins across many cycles of growth and division. Most proteins in bacteria are highly expressed, with a relatively small effect of number fluctuations (30) and with degradation negligible over the timescale of a cell-cycle (31, 32). These properties result in protein content being a global cellular variable, buffered from many microscopic noise sources. Its global nature results in a universal distribution shape, insensitive to many microscopic control parameters (33). Previous work has shown that the long-term dynamics of protein accumulates exponentially within the cell cycle and exhibits effective homeostasis similar to cell size; thus, it can be described by the same mapping model applied to cell size (7, 12). This analogy suggests a strong coupling between cell size and protein.

To further characterize this coupling, we have measured cell size simultaneously with two highly expressed proteins (metabolically relevant and irrelevant) in the same cell, and analyzed how these components of the phenotype co-evolve over multiple generations. All components accumulate exponentially, exhibiting a strong positive correlation between the accumulation rates on a cycle-by-cycle basis. Nevertheless, the measured relation between them is inconsistent with a simple proportion or a dominance of one component (e.g. cell size) that determines all others up to some noise. These results suggest treating proteins and cell size as coupled components of a multi-dimensional interacting system. Therefore, we studied such models with different coupling schemes between the components, and compared them to the integrated set of experimental results.

At the level of mapping between generations with effective feedback, it is difficult to achieve homeostasis of multiple components that exponentially accumulate and divide when one variable controls division. Perhaps surprisingly, this is true even if their exponential accumulation rates are identical up to a reasonable noise level. We found that a simple way to induce such homeostasis is by including the dynamic coupling between components during the cell cycle. Random linear interactions were sufficient to produce effective auto-catalytic dynamics during the cell cycle (26, 27); we have used this simple model despite the known non-linearity inherent to metabolic reactions. Cell division control was described by an effective restraining force. The finite duration of cell cycle, the small dynamic range of exponential accumulation ($\sim \times 2$), and the imperfect nature of division cause reshuffling of the different phenotype components at division. As a consequence, rather than a pure exponential accumulation, all components accumulate with effective exponents, which vary over cycles and between components, while maintaining a positive correlation among them.

This model induces simultaneous homeostasis on all cellular components, although only one may actually affect cell division. Moreover, it results in all components exhibiting negatively correlated accumulation and initial value. This correlation is manifested as an effective restraining force, whose specific value can vary depending on interactions and levels of noise. Consequently, different individual cells may have variable empirically measured values of this homeostasis parameter, including those measured for cell size.

*Predicted consequence of variable homeostasis: an independent verification.* The variable value of the homeostasis parameter is reflected in the structure of the auto-correlation function for cell size and protein content. While averaging over the noise ensemble smooths out such oscillations, the period of individual traces can be predicted by a theory applied to the mapping model across generations (28). This prediction depends explicitly on $\tilde{\beta}^{(k)}$ and agrees with the data, providing additional, independent, support for our finding of effective homeostasis parameters $\tilde{\beta}^{(k)}$ which vary across traces ($k$) of individual cells.

*Implications to division regulation.* Our results taken together highlight several gaps in the current understanding of cellular homeostasis in bacteria. Why are homeostasis parameters different among cells? We have provided an illustrative model consistent with the results, where one variable controls division, and measurable correlations are indirectly induced or modulated by intracellular interactions and noise. However, there could be other sources; for example, the micro-environment in the trap could affect processes in the cell that would result in such variability. Our analysis highlights the elusive nature of the homeostasis parameter $\beta$, and the difficulty in identifying what it represents in terms of intracellular processes, and in particular its relation to cell-division regulation. One possibility is that division regulation is an emergent property of the cell, which arises dynamically from complex interactions. Such dynamic feedback has been suggested as an organizing principle for mesoscopic-scale systems (25). Supporting this notion is our observation that cell size is controlled to a narrow region around distinct values for each trace, and the possible role of division fraction in homeostasis which has not been investigated so far. A combination of cell-size accumulation and division fraction as relevant control variables would certainly imply a global and integrated mechanism. It is also possible that some composite variables influence division more than any one of those currently measurable (34). These speculative possibilities remain to be investigated in future work.

## Methods

**Experimental procedure and data processing**

Wild type MG1655 *E. coli* bacteria were used in all experiments. Protein content was measured through the fluorescence intensity of green fluorescent protein (GFP) or red fluorescent protein (tdTomato) inserted into the bacteria on a high or medium copy number plasmid and expressed under the control of the promoter of interest. For measuring the expression level of a metabolically relevant protein, GFP was expressed from the medium copy number plasmid pZA (35) under the control of the *lac* promoter. For a metabolically irrelevant protein, GFP was expressed from the same plasmid pZA but under the control of the viral λ-phage PR promoter. For simultaneous measurement of the expression of two proteins, GFP was expressed from the high copy number plasmid pUC19 under the control of the *lac* promoter, while tdTomato was expressed from the pZA plasmid under the control of the λ-phage PR promoter.

Cultures were grown overnight at 30°C, LB medium (most cell size data in the main text and protein expressed from λ-PR promoter), or in M9 minimal medium supplemented with 1g/l casamino acids and 4g/l lactose (M9CL; cell size data presented in SI-4, protein expressed from the *lac* promoter, and simultaneous measurements of cell size and expression from both promoters, Fig. 4). The following day, cells were diluted in the same medium and regrown to early exponential phase, Optical Density (OD) between 0.1 and 0.2. When reaching the desired OD, cells were concentrated 10X into fresh medium and loaded into a microfluidic trapping device (see SI-9-1). After trapping, fresh medium was flown continuously through it to supply nutrients.

Cells were allowed to grow in the device for dozens of generations while maintaining the temperature fixed, using a made-in-house incubator. Images of the channels were acquired every 3 to 6 minutes in phase contrast and fluorescence modes using a Zeiss Axio Observer microscope with a 100x objective. The size and fluorescence of the tracked mother cell were measured from these images using the image analysis software microbeTracker (36). These data were then used to generate traces, such as those presented in Fig. 1A (see also Fig. SI-2-2). Growth stability of cells in the microfluidic device was verified by comparing the average division time in the first and second halves of the trace. No trend was detected in any of the experiments (see SI-9-2).

**Data analysis**

Single-cell traces were analyzed using home-made MATLAB programs. Trace autocorrelation functions and linear curve fitting were calculated by their implementations in MATLAB toolboxes. Homeostasis parameters $\beta$ (both for cell size and for protein) were estimated as the slope of the best linear fit to scatter plots, such as Fig. 1B, namely, exponential accumulation as a function of log cell size (or protein) at the start of each cycle. Using this fit, $\xi_n$ was estimated as the difference between the data and the line. The standard error of the slope in the linear fit for a single trace was estimated as $\sigma_{in}^2(\hat{\beta}) = \frac{\sigma^2(\xi_n)}{S^2}$, where $S^2 = \sum_{n=1}^{N}(\ln x_n - \overline{\ln x})^2$, $N$ the number of cycles in the trace, with the trace index $(k)$ omitted here for clarity. Data measured at the Jun lab were extracted from the webpage accompanying (3), and analyzed in the same way as our data (see SI-10).

The temporal average over a particular trace $k$ is represented by an overbar, e.g. $\overline{\ln x^{(k)}} = \frac{1}{N_k}\sum_{n=1}^{N_k} \ln x_n^{(k)}$, where $N_k$ is the number of cycles in the trace and $x_n^{(k)}$ measurements in that trace. The internal variance over the trace is computed as: $\sigma_{in}^2(\ln x^{(k)}) = \frac{1}{N_k}\sum_{n=1}^{N_k}\left(\ln x_n^{(k)} - \overline{\ln x^{(k)}}\right)^2$. The corresponding standard deviation $\sigma_{in}(\ln x^{(k)})$ is the square root of this quantity. To characterize the entire set of traces, the average across traces is denoted by brackets: $\langle\sigma_{in}(\ln x)\rangle = \frac{1}{M}\sum_{k=1}^{M}\sigma_{in}(\ln x^{(k)})$, with $M$ the number of traces.

The external variance quantifies the spread of temporal averages among traces and is computed as: $\sigma_{ex}^2(\overline{\ln x}) = \frac{1}{M}\sum_{k=1}^{M}\left(\overline{\ln x^{(k)}} - \langle\overline{\ln x}\rangle\right)^2$. We note that the difference between averaging the logarithm and taking the log of average quantities was not significant in our data. For initial cell size, this amounted to a 2.5% discrepancy, while for division fraction it was less than 1%. For the analysis of Fig. 3, the mapping model for individual traces (Eq. (4)) was simulated with parameters mimicking the measured data: number of traces, number of cycles in each trace, etc. Internal and external variances were computed similarly for measured and simulated traces.

**Model simulation**

We simulated the multi-component phenotype model Eqs. (4,5), for a vector of $N = 50$ components. Interactions $K_{ij}$ were independently drawn from a Gaussian distribution with mean $\frac{1}{\sqrt{N}}$ and standard deviation $\frac{1}{\sqrt{N}}$. This matrix was kept fixed over the entire ensemble of simulated traces, each with a number of cycles drawn uniformly between 30 and 250. The "nominal" homeostasis parameter was taken to be $\beta = 0.5$, similar to the value of the ensemble average of the experimental data. The common pivot-point coordinates are taken as $x^* = 3$ and $\phi^* = \ln 2$. The end of each cell-cycle $n$ is determined by Eq. (2), applied to the first component $x_n^{(1)}$. At division, each component $j$ is multiplied by an independent Gaussian variable $f_{j,n}$ with mean 0.5 and standard deviation $\sigma_f = 0.1$, truncated to (0,1). The noise in division control, $\xi_n$, is a zero-mean Gaussian variable with standard deviation $\sigma_\xi = 0.2$.

**Figure 1: Correlations in cycles of exponential accumulation and division.**
(A) A portion of a trace measuring the size of a trapped bacterium along time, illustrating the exponential accumulation within cycle $n$, $e^{\phi_n}$, and the division fraction, $f_n$. These two variables connect the initial cell size $x_n$ with that at the next cycle $x_{n+1}$ (see Eq. (1)); both fluctuate from one cycle to the next. (B) Exponential accumulation $\phi_n$ is negatively correlated with $\ln x_n$ (best fit slope for Eq. (2): $\beta = 0.49 \pm 0.02$). Taking $x^*$ to be the average cell size, $x^* = 2.7 \mu m$, we find $\phi^* = 0.69 \pm 0.2$. (C) Blue line: a long sequence of initial cell sizes $x_n$, from one trace as a function of cycle number. Grey line: a shuffled process created from the measured pairs $(e^{\phi_n}, f_n)$, by applying them as fold changes to the initial condition of the trace in random order, thus discarding the correlation between fold-change and initial cell size.

**Box 1: Possible patterns of variability and invariants in single-cell trace correlations**
(A) Single cell traces could exhibit the same slope of correlation, indicating that the effective restraining force strength is a relevant control variable. (B) Alternatively, they could exhibit an invariant intersection point, pointing to a preferred common cell size. (C) Traces could also be variable in both properties without conserving any global invariant. In all panels, grey points are measurement data. Black dots result from simulating the mapping model (Eqs. (1,2)), with different slopes and intercepts $(\beta, \phi^*)$ for the correlation of Eq. (2). In all cases the average division ratio is ½ and the average fold-change over the trace is 2.

**Figure 2: Individual (A,B) and common (C,D) aspects of cell-size homeostasis in bacterial traces**
(A) The same data as in Fig. 1B are plotted in gray for the entire population. Points from two individual traces are highlighted in color, with their respective best linear fits, displaying a different slope for each and thus a different homeostasis parameter $\beta$. (B) Estimated slopes $\beta$ for all individual traces as a function of their length in number of generations, with error-bars denoting the standard error in the estimate (see Methods). (C) Best linear fits for all individual traces intersect at a common pivot-point. Green circle: ensemble average of the two axes. (D) A flow map is estimated in the two-dimensional phase space $(\ln x_n/x^*, \phi_n)$. The flow direction is indicated by arrows; its amplitude is encoded in the underlying heat map (contours have a uniform spacing of 0.1, and range from 0 [black] to 2 [white]). The pivot point of Fig. 2C (green circle) is an attractor on this projection of the dynamics.

**Figure 3: Individuality and slow dynamics in cell size traces**
(A) Two measured individual traces (colors) showing cell size in consecutive cell-cycle starts, $x_n$ (3-pt smoothed). Horizontal lines: time-average of each trace. Right: probability density functions (pdf) of cell size values for the two plotted traces (corresponding colors). Each cell maintains fluctuations around a distinct mean value, with internal standard deviation $\langle \sigma_{in}(\ln x) \rangle = 0.26$ averaged over all traces. The standard deviation of this quantity among traces is 0.06. (B) The same as in A, for two simulated traces (Eq. (4)), with parameters matching those in A. The internal standard deviation of simulated traces, $\langle \sigma_{in}(\ln x) \rangle = 0.43$, is larger for model traces. The standard deviation of this quantity across traces is $\langle \sigma_{ex}(\overline{\ln(x/x^*)}) \rangle = 0.05$. (C) Two measured traces exhibit distinct clustering in the $(\ln x_n, \phi_n)$ plane (colors). As a consequence, each cell maintains a distinct average size (large circles) over its lifetime. (D) Temporal averages of division ratio and accumulation exponents in all measured traces. Solid line: $\overline{\phi_n} + \overline{\ln f_n} = 0$. The two traces shown in (C) are highlighted in color.

**Figure 4: Relationship between cell size and content of highly expressed proteins**
(A) Traces of cell size (top) and two highly expressed proteins (fluorescent proteins expressed from the *lac* promoter, middle; and the λ-phage PR promoter, bottom). All three components exhibit cycles of exponential accumulation and division. (B) Exponential accumulation rates of cell-size ($\alpha_S$), *lac* expression ($\alpha_L$) and λ-PR expression ($\alpha_\lambda$) are strongly correlated across cycles. Each dot in the 3D space represents the three exponential rates corresponding to one cell cycle. (C) Protein density in three individual cells (different colors), each collected over multiple cycles of growth and division, displays a broad distribution.

**Figure 5: Model of interacting cellular components**
(A) Traces of three components out of 50, which interact linearly within the cell-cycle according to a random interaction matrix. One component (top trace) controls cell division through Eq. (2) with $\beta_1 = 0.5$, while the other components follow and segregate their content randomly at division time. Accumulations within the cell-cycle reflect global dynamics of all components and are given by a combination of exponentials which can be described to an excellent approximation by an effective exponent (see SI-7). Both the controlling component (B) and the non-controlling components (one example shown in (C)) exhibit effective homeostasis, namely, a negative correlation between the component at cell-cycle start and its exponential accumulation along that cycle (150 traces simulated, each consisting of $30 - 250$ division cycles). (D) Effective homeostasis depends on model parameters: heat-map of the empirically estimated homeostasis parameter for the controlling component, $\tilde{\beta}_1$, averaged over 100 model realizations, as a function of noise parameters. The interaction matrix, division control, and system size are kept fixed across realizations. (E) Experimental values of homeostasis parameters $\tilde{\beta}_j$ estimated from all traces are plotted in color (red, cell size; green, protein content) as a function of the noise level $\sigma_\xi$ (see Eq. 2 and Methods). Model prediction with $\sigma_f$ fixed at 0.2 (horizontal projection of (D)) is depicted by a black line.

**Figure 6: Auto-correlation functions (ACF) of individual traces**
(A) ACFs computed from several cell-size traces (colors). Black: average over all traces. (B) ACF for traces of fluorescent protein expressed from the λ-PR promoter. Black: average over all traces. (C) Mean peak-to-peak distance $M(\beta)$ computed from all individual traces of cell size and $\lambda$-PR expression. Binned data is shown in large circles. Black curve: prediction from theory (28; see SI-8).

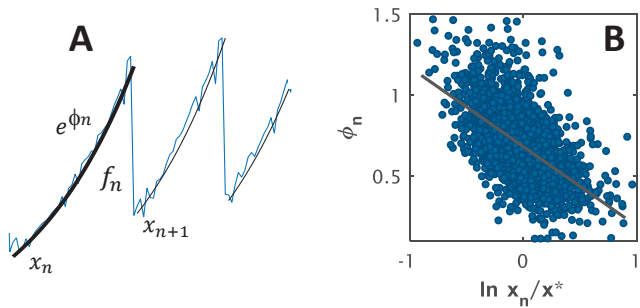
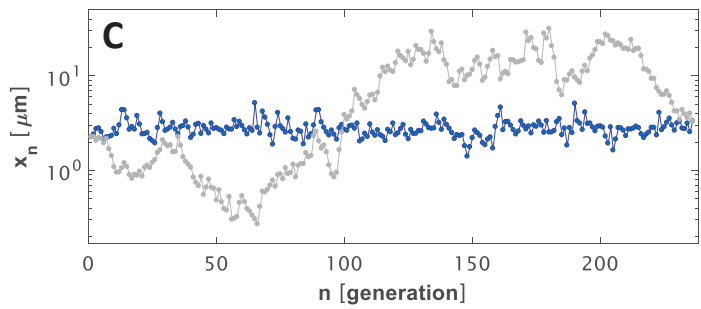

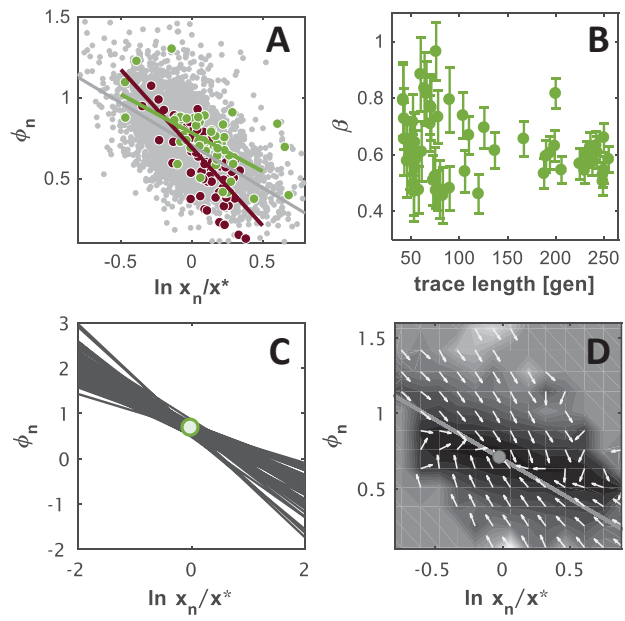

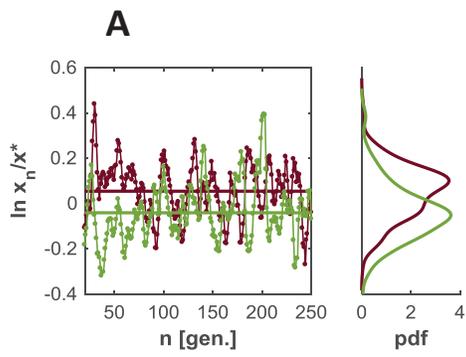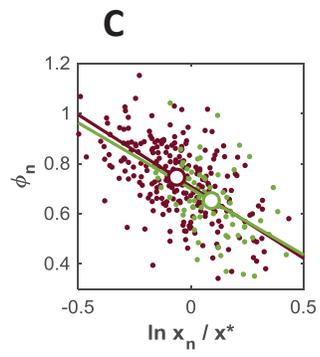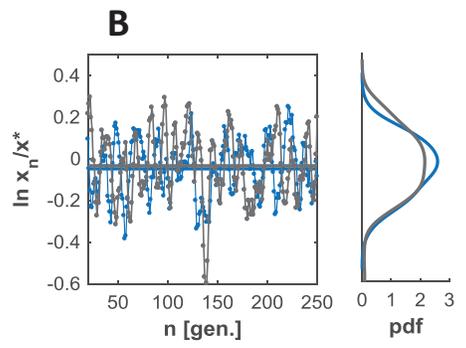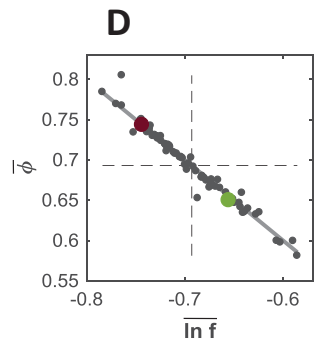

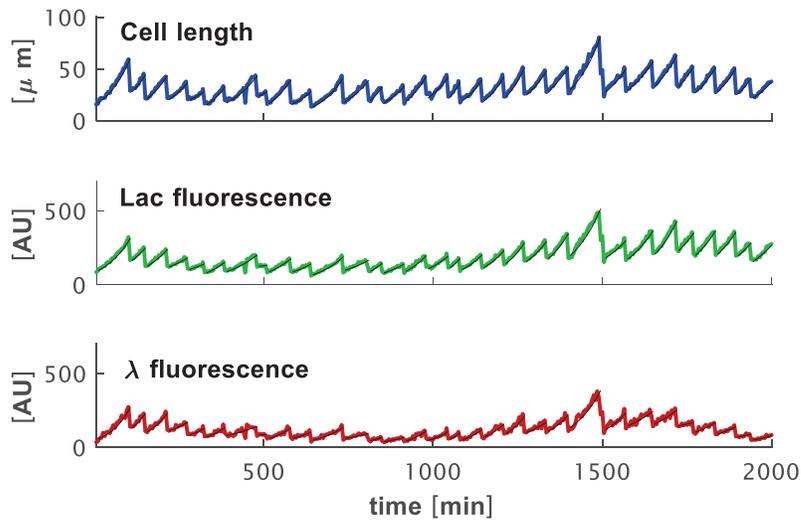
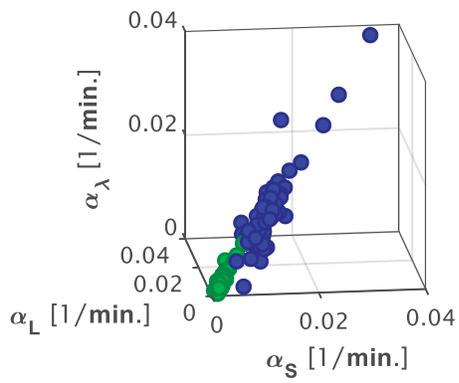
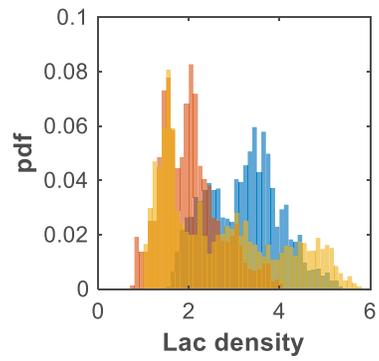

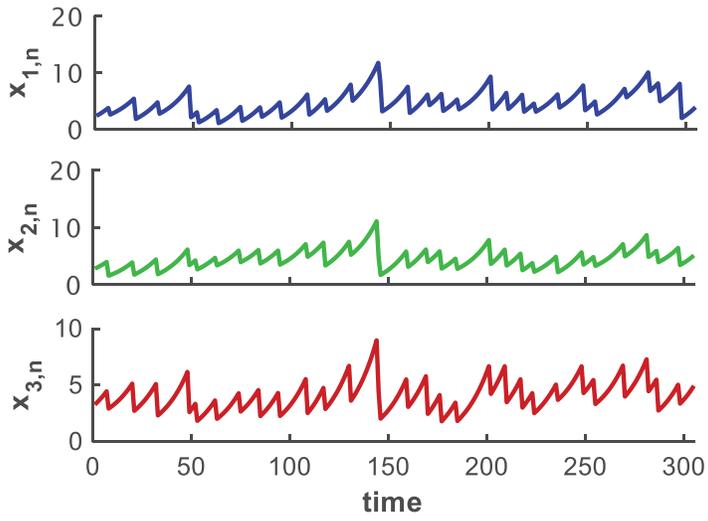
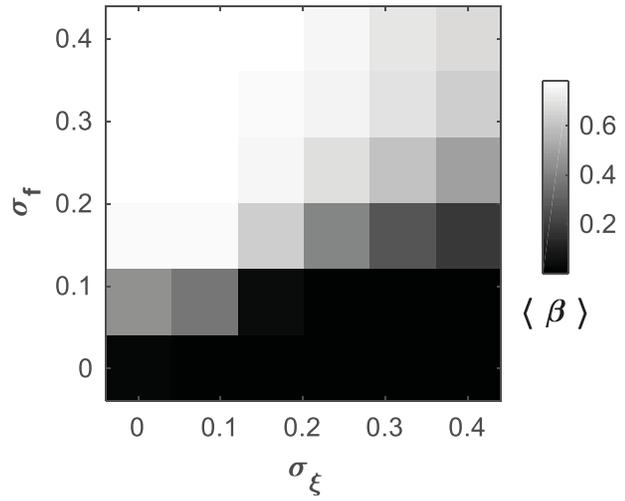
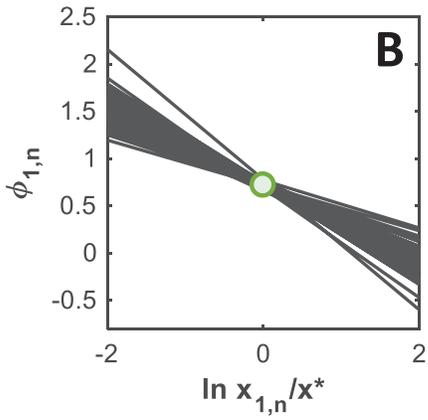
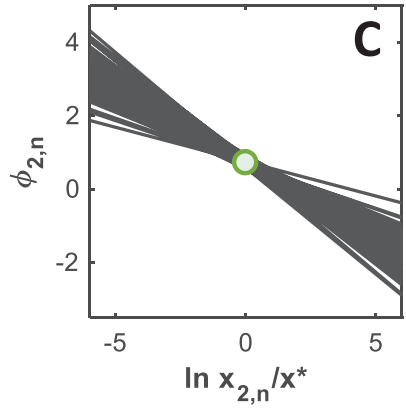
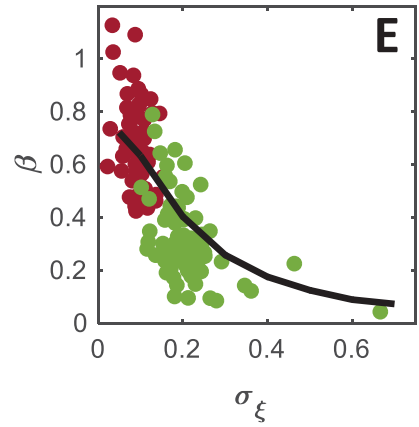

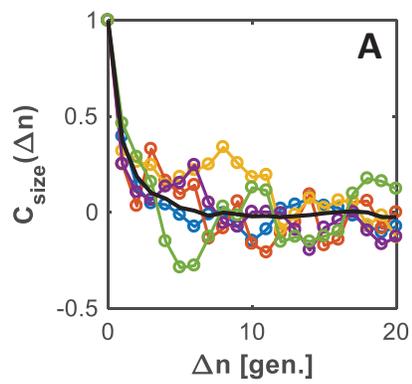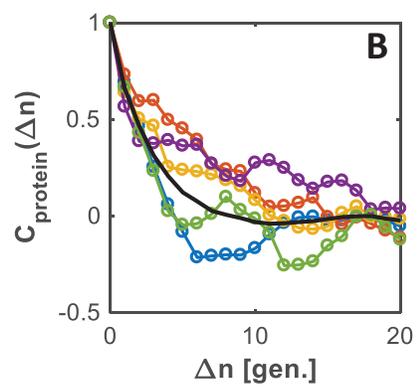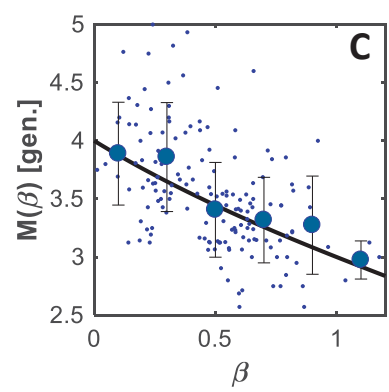

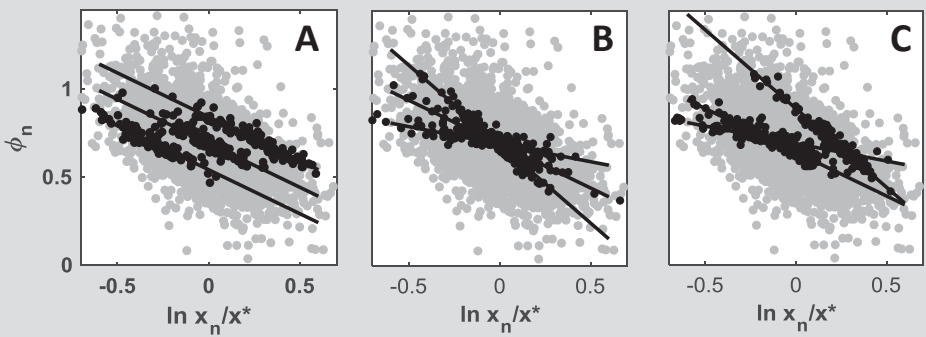

**Box 1: Possible patterns of variability and invariants in single-cell trace correlations**
(A) Single cell traces could exhibit the same slope of correlation, indicating that the effective restraining force strength is a relevant control variable. (B) Alternatively, they could exhibit an invariant intersection point, pointing to a preferred common cell size. (C) Traces could also be variable in both properties without conserving any global invariant. In all panels, grey points are measurement data. Black dots result from simulating the mapping model (Eqs. (1,2)), with different slopes and intercepts ($\beta, \phi^*$) for the correlation of Eq. (2). In all cases the average division ratio is ½ and the average fold-change over the trace is 2.

# Supporting Information: Individuality and slow dynamics in bacterial growth homeostasis

## SI-1: Additional statistics over all traces

In this section we present, for completeness, additional global statistical. All panels are created from the entire data-set in LB medium, which includes 8152 cell cycles pooled from 79 single-cell traces. Note that all statistics plotted here are model-independent and rely only on the parameterization of the data illustrated below.

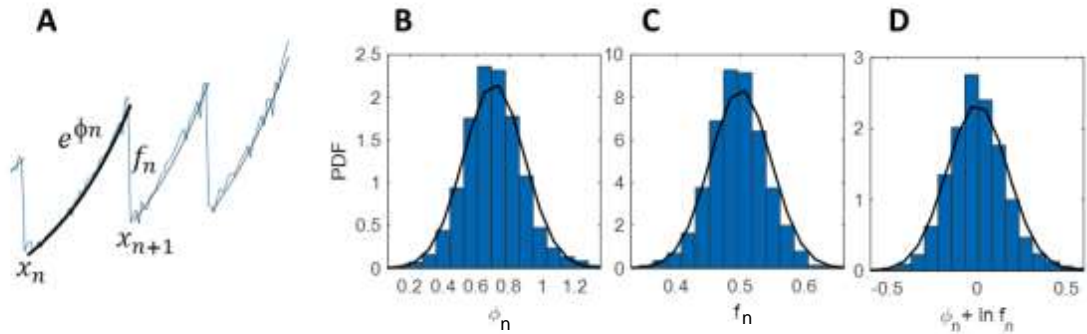

**Figure SI-1-1: Distributions of single-cycle parameters.** (A) Parameterization for individual cell-size measurement trace. (B) Distribution of accumulation exponent over the cycle, mean 0.707, std 0.18 (about 26%. Note that ln2 = 0.693). (C) Distribution of division fraction across the entire data-set. Mean 0.495, std 0.047 (about 10%). (D) The distribution of total logarithmic change across a cycle, $\phi_n + \ln f_n$, is symmetric with an average of $10^{-4}$ and a standard deviation of 0.17.

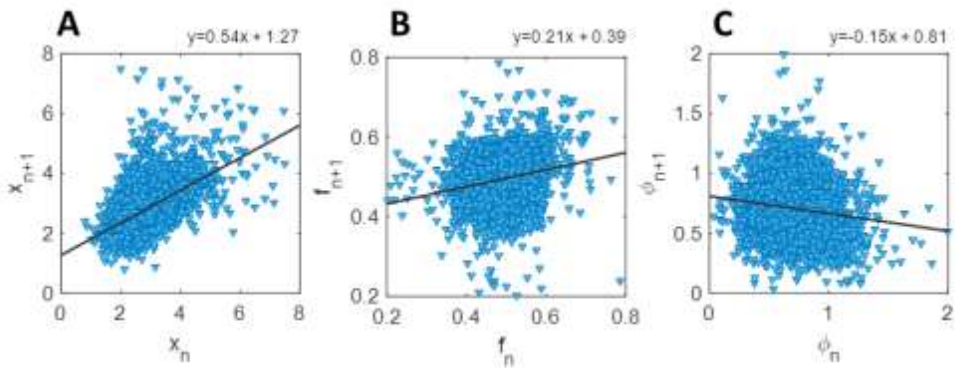

**Figure SI-1-2: Correlations across consecutive cycles.** Empirical correlations between consecutive generations in all dynamic parameters: cell size (A), division fraction (B), and accumulation exponent (C). Best linear fits are displayed on the figures.

## SI-2: Model equivalence in special cases

We show that our mapping model for size across cell-cycles, which was constructed empirically from the data, reduces to the model of Amir (1) under simplifying assumptions and in the linear approximation.

We start from the description of the trace as exponential accumulations and divisions,

$$x_{n+1} = f_n x_n e^{\phi_n}, \qquad \text{(Eq. SI-2-1)}$$

and the correlation of the accumulation exponent with initial size:

$$\phi_n = \phi^* - \beta \ln \frac{x_n}{x^*}, \qquad \text{(Eq. SI-2-2)}$$

where we have neglected the noise for this derivation. Now assume that the division ratio $f_n$ is exactly 1/2 in each cell cycle, and that the average accumulation satisfies $\phi^* = \ln 2$, and insert back into (SI-2-1):

$$x_{n+1} = \frac{1}{2} x_n 2 \, e^{-\beta \ln \frac{x_n}{x^*}} = x_n^{1-\beta} (x^*)^\beta. \qquad \text{(Eq. SI-2-3)}$$

Denoting the size at division and birth by $x_d$ and $x_b$ respectively, (SI-2-3) implies

$$x_d = 2 x_b^{1-\beta} (x^*)^\beta,$$

which is the relation suggested by Amir (1). Now, expand this relation to first order in the deviation of $x_b$ from the ensemble average $x^*$:

$$x_d \sim 2(x^*)^\beta \left[ (x^*)^{(1-\beta)} + (1-\beta)(x^*)^{(-\beta)}(x_b - x^*) \right] = 2(1-\beta)x_b + 2\beta x^*.$$

This approximate relation between size at birth and size at division has three special cases:

$$\begin{aligned}
\beta &= 0 & x_d &= 2 x_b \\
\beta &= 0.5 & x_d &= x_b + x^* \\
\beta &= 1 & x_d &= 2 x^*
\end{aligned}$$

which can be intuitively interpreted as "timer" control (doubling of cell size during the cycle, equivalent to a constant time under the assumption of constant exponential rate); "adder" control, where a constant mass is added each cycle; and "sizer" control, where the cell needs to reach a threshold size to divide.

To further connect to previous work (1, 2), we present here the analysis of our data in terms of added cell-size per cycle, and size at the cycle end, as a function of initial size (Fig. SI-2-1, top and bottom rows respectively). Over the entire collection of traces, the data exhibits no significant correlation between added size, $\Delta x$, and initial cell size (panel A top; note however that the slope is different from zero within the estimate error). Correspondingly, the slope of final vs. initial size is close to 1 (A bottom). The error estimate of these slopes is reduced by the large statistics but increased by a large spread of the data around the fit. When the same analysis is applied to long and stable traces (B-D), it is found that the ensemble average correlation is in fact composed of a mixture of different slopes for individual cells. The added size and final size give expected corresponding slopes (the difference between them is 1).

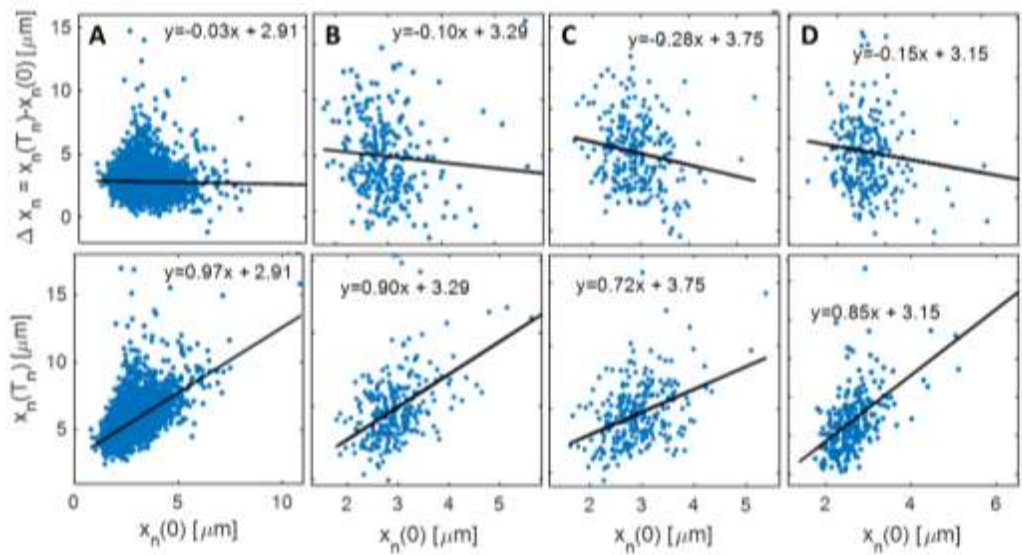

**Figure SI-2-1. Alternative approaches to size analysis.** The cell size (represented by length in µm) added during each cell-cycle (top panels) and cell size at the end of the cycle (bottom panels) are plotted as a function of the cell size at the start of the same cycle. (A) Analysis of the entire data-set. Standard deviation of noise around the fit: 1.07. Due to the large sample obtained from pooling all traces (~8000 cycles), the uncertainty in the slope estimate reduces to $\pm\ 0.02$. (B-D) Three of the longest traces in our data can be analyzed in a similar manner. (B) Trace of length $N = 255$, slope = $-0.1 \pm 0.09$; (C) $N = 250$, slope = $-0.28 \pm 0.1$; (D) $N = 249$, slope = $-0.15 \pm 0.1$. The two fits (top and bottom panels for each trace) are obviously related to one another for each individual trace, and the standard error – the same.

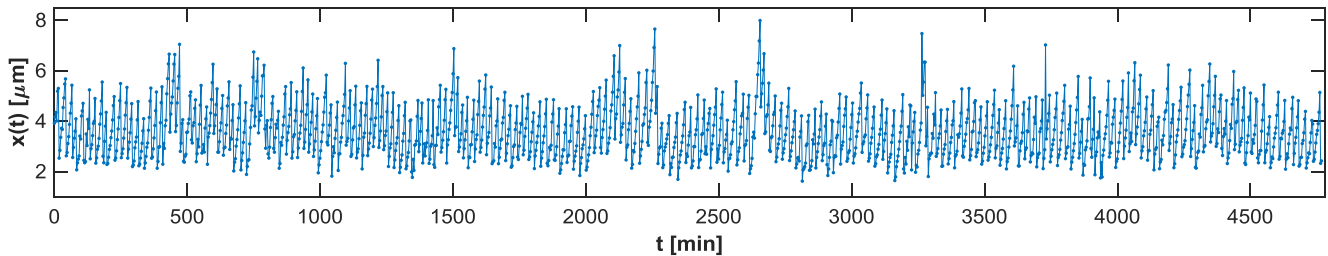

**Figure SI-2-2.** Cell size vs. time for one of the long traces, demonstrating its stability.

## SI-3: Statistical analysis of slopes in individual bacterial traces

To assess individuality of different bacterial traces in terms of their homeostasis parameter $\beta^{(k)}$, we use two types of analysis: a measure of ergodicity breaking used in statistical mechanics (3), and a statistical test of the hypothesis that the two estimates of the linear slopes are drawn from the same process (4).

Assume first that all individual traces are statistically equivalent and drawn from a well-defined stochastic process represented by Eq. (3) in the main text with fixed parameters. Our best estimate of these parameters is obtained by using the whole data set: we use the correlation between $\phi_n$ and $\ln x_n$ over all ~8000 cell cycles to estimate it as

$$\phi_n = 1.23 - 0.49 \ln x_n + \xi_n \qquad \text{(Eq. SI-3-1)}$$

(see Fig. 1(B) in main text), with $\xi_n$ the measured noise around the linear fit. This is found to be, to an excellent approximation, a Gaussian with zero mean and standard deviation $\sigma(\xi_n) = 0.16$.

Now we simulate from this process a trace of exponential accumulations and division, where division is distributed randomly around ½ and accumulation is drawn as a function of current size from (Eq. SI-3-1); we construct the same number of traces and trace lengths as those measured in our experiment. For the simulated traces, we estimate the correlation slopes in the same way as was done for the data. This procedure gives a benchmark for how different from one another the individual estimated slopes are expected to be, as a result of noise and finite sampling. We find that the variance among traces in the experiment is more than twice (0.025) the variance among simulated traces (0.012).

To further quantify the set of estimated slopes, an "ergodicity breaking" parameter $EB$ can be defined which measures to what extent the difference among the traces stems from variability of the entire ensemble, as opposed to a real underlying distinction between traces. The best estimate of the slope is obtained by the following average over the k-th trace

$$\hat{\beta}^{(k)} = \frac{\sum_{n=1}^{N_k} \left( \ln x_n^{(k)} - \overline{\ln x^{(k)}} \right) \phi_n^{(k)}}{S_k^2},$$

with $S_k^2 = \sum_{n=1}^{N_k} \left( \ln x_n^{(k)} - \overline{\ln x^{(k)}} \right)^2$. The external variance of $\hat{\beta}^{(k)}$ estimates across the collection of $M$ traces is then

$$\sigma_{\text{ex}}^2(\hat{\beta}) = \frac{1}{M}\sum_{k=1}^{M}(\hat{\beta}^{(k)})^2 - \left(\frac{1}{M}\sum_{k=1}^{M}\hat{\beta}^{(k)}\right)^2.$$

This is a measure of how different the estimated slopes are from one trace to the other. This variance needs to be compared to the inaccuracy of the estimate in one trace due to the finite sample and the noise, which is given by

$$\sigma_{\text{in}}^2(\hat{\beta}^{(k)}) = \frac{\sigma^2(\xi_n)}{S_k^2},$$

(4). For a fixed noise level, $\sigma_{\text{in}}^2(\hat{\beta}^{(k)})$ decreases to zero as the size of the sample increases ($S^2 \propto N_k$). The Ergodicity Breaking parameter is defined as (3):

$$EB(\hat{\beta}) = \frac{\sigma_{\text{ex}}^2(\hat{\beta}) - \langle\sigma_{\text{in}}^2(\hat{\beta})\rangle}{\sigma_{\text{ex}}^2(\hat{\beta})},$$

where we have used the notation ⟨ ⟩ for averaging over traces. If the variation between traces is within the error of each trace, then we will have approximately $EB = 0$. If, however, the variance among trace estimates is larger than the error in each trace, we will have a nonzero $EB$. In our data we find $EB = 0.49$ for the measured data; for the simulated traces shown in the figure, we find $EB = 0.04$ (repeating the simulation results in slightly different values depending on realization, but these are invariably much smaller than the data). This suggests that the distinction between traces is beyond statistical noise.

We apply also a statistical test for the hypothesis that two traces are consistent with the same linear correlation; this takes into account explicitly the noise in the correlation and the number of samples. In this approach, we consider the estimates $\hat{\beta}$ as random variables. For a sampling of length $N_k$ from the uniform process in (Eq. SI-3-1), this random variable has average $\beta$ and variance given by $\sigma_{in}^2(\hat{\beta}^{(k)})$ defined above.

To address the hypothesis that two such estimates for two different traces are random variables drawn from one and the same distribution, one defines the normalized difference variable, $d_{kl} = (\beta_k - \beta_l) \big/ \sqrt{\sigma_{in}^2(\hat{\beta}_k) + \sigma_{in}^2(\hat{\beta}_l)}$. Estimation theory predicts that, for samples drawn from a common process and with Gaussian noise, $d_{kl}$ is a standard normal variable (average zero and variance 1) (4). Indeed, for the simulated traces sampled from the global process, Fig. SI-3-1B shows this variable is normal – with an average of zero and a standard deviation of 1 (light gray histogram). In contrast, the experimental data exhibit a zero mean but a standard deviation of 1.41, more than 40% above the theoretical predictions. For the collection of traces in our experiment, the histogram is constructed from hundreds of pairs, and this large sampling would suggest a precise fit to estimation theory, as indeed is found for the synthetic traces. This result refutes the hypothesis, that all traces were drawn from the same ensemble and that the differences between slopes is explained by random noise and finite sampling.

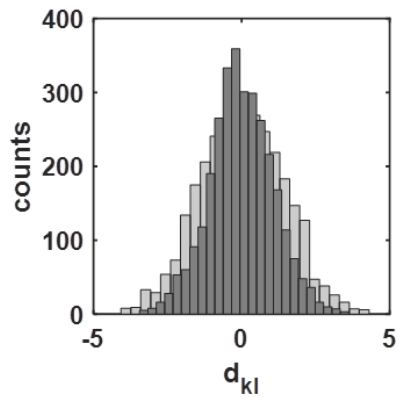

**Figure SI-3: Statistical analysis of distinct traces.** The difference between individual slopes, normalized appropriately by the standard error and the sample size (see text for details), should be a normal Gaussian distribution. This is true for the simulated traces (dark gray) but not for the real data, which has a standard deviation of 1.4.

## SI-4: Cell size homeostasis in a different medium

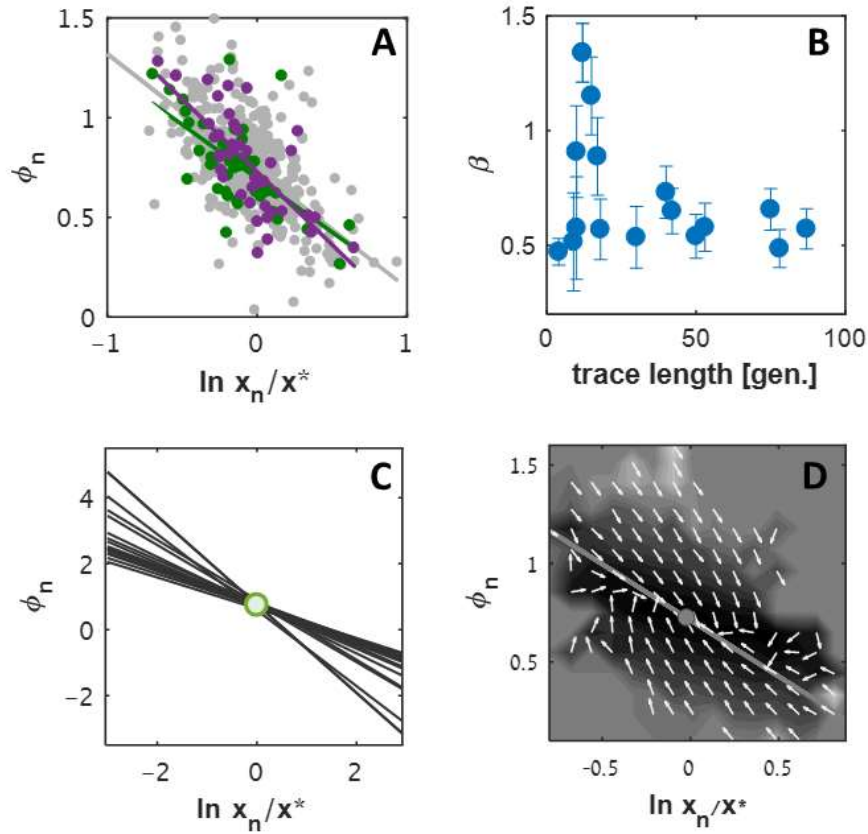

**Figure SI-4. Individual (A,B) and common (C,D) aspects of cell-size homeostasis in M9CG medium.** Same graphs as in Fig. 2 in the main text, for data collected in M9CG medium instead of LB (17 individual traces). (A) Cell size accumulation exponents plotted in gray for the entire data-set. Points from two individual traces are highlighted in color, with their respective best linear fits, displaying a different slope for each and thus a different homeostasis parameter $\beta^{(k)}$. (B) Estimated $\beta^{(k)}$ for all individual traces as a function of their length in number of generations, with error-bars denoting the standard error in the estimate. (C) Best linear fits for all individual traces display distinct slopes, and intersect at a pivot-point common to all cells. Green circle: average of axes (cell size and accumulation exponent) across the entire data-set. (D) A flow map is estimated in the two-dimensional plane $(\ln x_n, \phi_n)$. The direction of the flow is indicated by the velocity arrows; the amplitude of the flow is encoded in the underlying heat map (grey levels have a uniform spacing of 0.1 ranging from 0 [black] to 2 [white]).

## SI-5: Effect of division ratio bias on time-averaged cell sizes of individual traces

Here we address the deviations between time-averaged cell size of individual traces within the framework of our model. We use the mapping model to compute the cell size in the two ensembles, i.e. averaging over the collection of traces and averaging over time in a single trace. Performing first the average over all cells we find:

$$\langle \ln x_n \rangle = \ln x^* + \frac{\phi^* + \langle \ln f_n \rangle}{\beta} = \ln x^*, \qquad \text{(Eq. SI-5-1)}$$

as expected, since over the entire data-set $\langle \ln f_n \rangle = \ln\left(\frac{1}{2}\right)$ whereas $\phi^* = \ln 2$. This makes explicit the independence of the steady-state on $\beta$. In contrast, when averaging over time in an individual trace, we find:

$$\overline{\ln x_n^{(k)}} = \ln x^* + \frac{\phi^* + \overline{\ln f_n^{(k)}}}{\beta^{(k)}}. \qquad \text{(Eq. SI-5-2)}$$

A deviation of the temporal average $\overline{\ln f_n^{(k)}}$ from the global average of $\ln\left(\frac{1}{2}\right)$ will create a mismatch weighted by the individual homeostasis parameter $\beta^{(k)}$, and induce a discrepancy between the temporal average and the common attractor $\ln x^*$. Indeed, we find that averaged over all traces in our data-set, the temporally averaged division ratio shows an external standard deviation of $\langle \sigma_{\text{ex}}(\overline{\ln f}) \rangle = 0.054$. For comparison, a collection of random fractions drawn from the ensemble-level distribution (with the same trace lengths as the data), is ~4.5 times smaller.

Fig. SI-5 displays the model prediction, Eq. (SI-5-2), with $\beta^{(k)}$ and $\overline{\ln f_n^{(k)}}$ estimated from individual traces, as a function of the discrepancy between temporal and global average for these traces. Since this discrepancy depends only on dynamic parameters of the trace and not on the average value itself, we plot results from two experiments in different conditions, showing a reasonably good agreement with the prediction. We note that the deviation from ensemble-average size due to slow dynamics of division fractions does not rely on variability in $\beta$; even two traces with the same $\beta$ could differ in their average size, depending on the temporal average $\overline{\ln f_n}$.

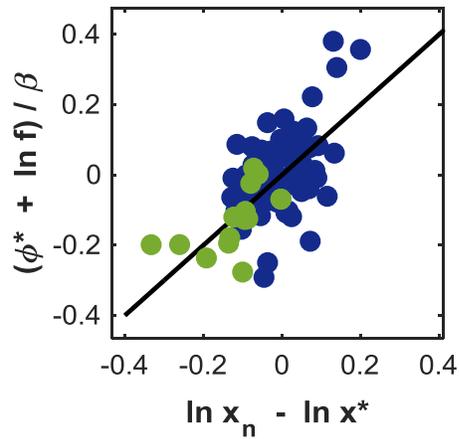

**Figure SI-5.** Deviations of temporally averaged cell size from ensemble average are computed directly from measurements (x-axis), and predicted from mapping model, Eq. (SI-5-2), using estimates of the parameters (y-axis). Blue circles: LB medium. Green circles: M9 medium.

## SI-6: Protein homeostasis

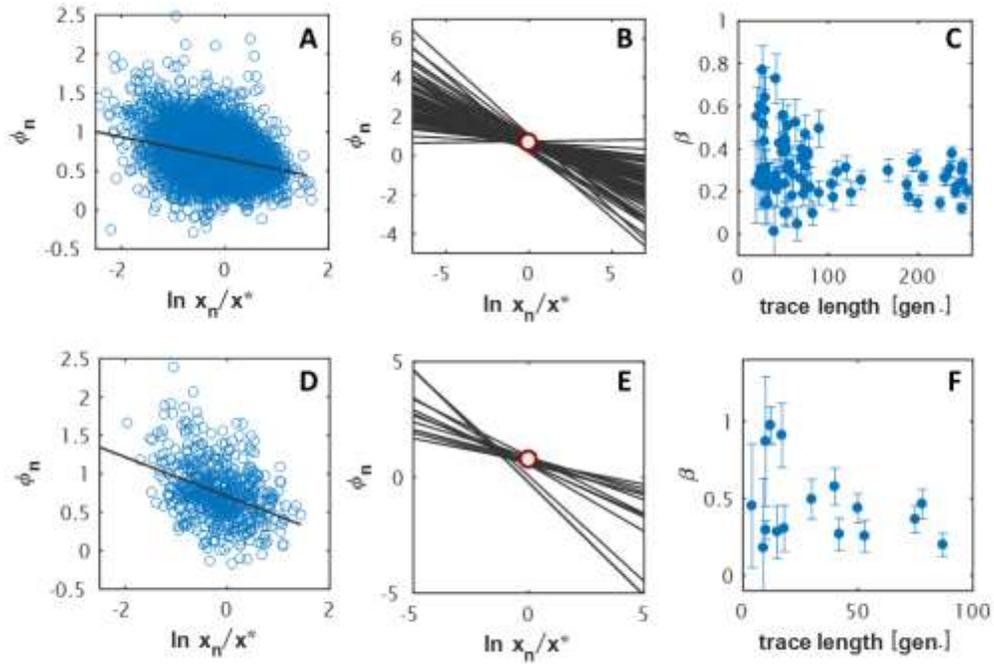

**Figure SI-6. Protein homeostasis: GFP expressed from λ-PR (top) and *lac* (bottom) promoters.** In general, highly expressed proteins display qualitatively similar dynamics as cell size, i.e. exponential accumulation and division, with negative correlation between exponential accumulation of protein during the cell-cycle and its initial value of at the start of the cycle. In these graphs $x_n$ represents the total fluorescence over the cell area, where GFP is expressed from two promoters: the λ-PR promoter (top panels) and the *lac* promoter (bottom). $x^*$ is the average fluorescence over the entire collection of cells. Scatter-plots (A,D) are shown for entire collection of cycles in the two experiments. Best fit lines have slopes of $\beta = 0.14$ (A), and $\beta = 0.26$ (D). (B,E) As for cell size, this global negative correlation is revealed as a superposition of distinct correlations for individual traces, each with a different slope, which intersect at the ensemble average. (C,F) Estimates of slopes, representing the protein homeostasis parameters $\beta$, with their error-bars as a function of trace length.

## SI-7: Multi-component phenotype model

### SI-7-A: Dynamic interactions model – results

Model description:

Intra-cycle dynamics $\quad\quad\quad \frac{d}{dt}\vec{x_n}(t) = \mathcal{K}\vec{x_n}(t), \quad 0 < t < T_n \quad\quad\quad$ (a)

Division conditions $\quad\quad\quad \phi_n = \phi^* - \beta \ln \frac{x_{1,n}(0)}{x^*} + \xi_n \quad\quad\quad$ (b)

Division fractions $\quad\quad\quad x_{j,n+1}(0) = f_{j,n} \cdot x_{j,n}(T_n), \quad\quad\quad$ (c)

where $\mathcal{K}$ is a random interaction matrix; $T_n$ the division time is determined when the accumulation exponent satisfies the condition in (b); and division in (c) describes any one of the components $j$. The noise variables in the model are both Gaussian with $\xi_n \sim \mathcal{N}(0, \sigma_\xi)$ and $f_{j,n} \sim \mathcal{N}\left(\frac{1}{2}, \sigma_f\right)$ (truncated to (0,1)). All noise variables are independent along time and across components. $x^1$ is the "controlling component" since the time to divide depends on the exponential accumulation in that variable.

The dynamics of all components during the cell-cycle can be directly calculated from this model. The resulting trajectory within cycle $n$ is given by:

$$\vec{x_n}(t) = \sum_{i=1}^{N} c^i e^{\lambda_i t} v_i, \quad\quad 0 < t < T_n. \quad\quad\quad \text{(Eq. SI-7-1)}$$

Here $\lambda_i$ are the eigenvalues of $\mathcal{K}$, $v_i$ are the respective eigenvectors, and $c^i$ are the projections of $\vec{x}_n(0)$ onto the basis of eigenvectors $V = [\vec{v_1} \ldots \vec{v_N}]$, i.e.

$$\vec{c} = V^{-1}\vec{x}_n(0). \quad\quad\quad \text{(Eq. SI-7-2)}$$

The resulting in-cycle trajectories are linear combinations of exponentials (Eq. SI-7-1), where at least some of them are positive. Such a combination can be described, to a good approximation over a finite time, by a single exponential with an effective accumulation rate. Fig. SI-7-1A (upper panel) illustrates a model trajectory of one phenotype component, fitted to effective exponential growth. Over long time-scales, a linear combination of exponential functions would be dominated by the leading exponent (5). However, biological constraints limit the cell-cycle to relatively short times, over which the components increase by only a factor of ~2. Due to this limited time the effective exponent depends on all eigenvalues as well as on their contribution to the initial condition at cycle start. Fluctuations across cycles in the effective exponent of any component are caused by the distribution of random fractions at the beginning of each cycle, which in turn reshuffles the prefactors of the exponents in Eq. SI-7-1.

Examples of the range of effective exponentials for one phenotype component along consecutive generations, all normalized to 1 at the cell-cycle start, are presented in Fig. SI-7-1A (lower panel). Model simulation results are plotted in blue dots whereas exponential fits are shown with black lines. Our previous work on traces of highly expressed proteins, has shown that variability in the exponential accumulation rates among generations is significant ($CV \approx 0.5$), and crucial for obtaining the broad universal protein distribution of individual traces (6). This variability arises here naturally from the effective interactions and division noise, without the need to explicitly introduce a large stochastic element into the rate. Moreover, even though the exponential accumulation rate of each phenotype component exhibits wide variability across cell-cycles, a strong correlation is still observed between the exponential accumulation rates of all components on a cycle-by cycle basis (Fig. SI-7-1B).

In Fig. 5 of the main text it was shown that all components acquire an induced negative correlation between their exponential accumulation and initial value, namely an effective homeostasis parameter $\beta$. Fig. SI-7-1C shows that different realizations induce different $\beta$ values of for the same component even when the matrix $\mathcal{K}$ is kept the same. Finally, as discussed in the main text, the variability in $\beta$ leads to distinct patterns in the autocorrelation function (ACF) of each component, which is washed away when all ACFs are averaged together (Fig. SI-7-1D).

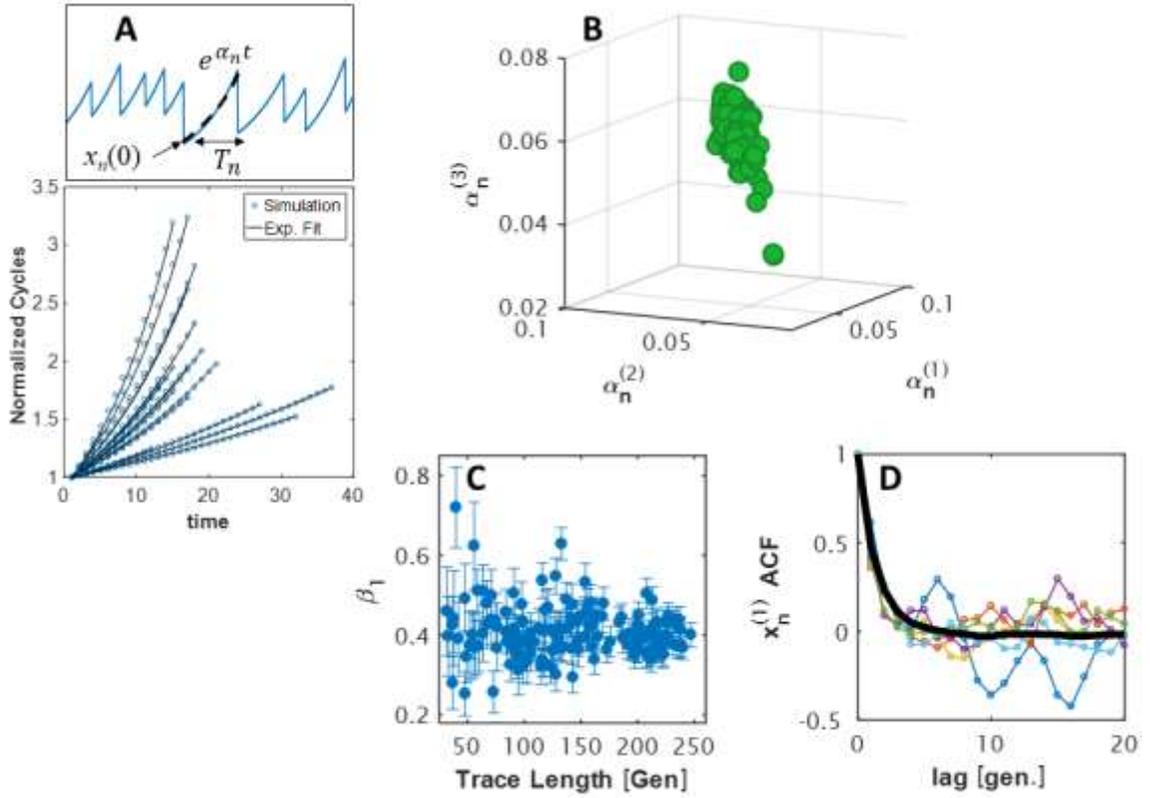

**Fig. SI-7-1. Simulation results of the multi-component model.** (A) In the upper panel, we present a section of a trace of one of the components generated in the model simulation. Normalizing each cell-cycle by the value of the component at the start of the cycle (lower panel) demonstrates that each cell-cycle in the trace exhibits a different effective exponential growth rate as seen in the experimental data. The exponential growth rates of different components, are correlated on the cycle-by-cycle basis as can be seen in (B) for three of the 50 components simulated. (C) Empirically estimated $\beta$ values for a total of 150 simulated traces, each consisting of $30 - 250$ division cycles. (D) ACF of different realizations are presented in different colors. The ensemble-averaged ACF (black line), exhibits simple exponential decay without any distinct pattern, similar to the experimental data.

### SI-7-B: Dynamic interactions are essential for homeostasis

The interactions among the different components during the cell-cycle are essential for stabilizing the exponential accumulation and division of all components over multiple cycles. When interactions are absent ($\mathcal{K}_{ij} = 0$ for $i \neq j$), stability will only be achieved for the controlling component, even if the growth rates of all components are identical ($\mathcal{K}_{ii} \equiv 1$). Stated differently, a picture where all proteins are enslaved to cell size and have the same exponential growth rates, is not only inconsistent with statistical properties of the data (Fig. 4 of main text), but is also theoretically inconsistent with global homeostasis in a multi-dimensional system over multiple cycles.

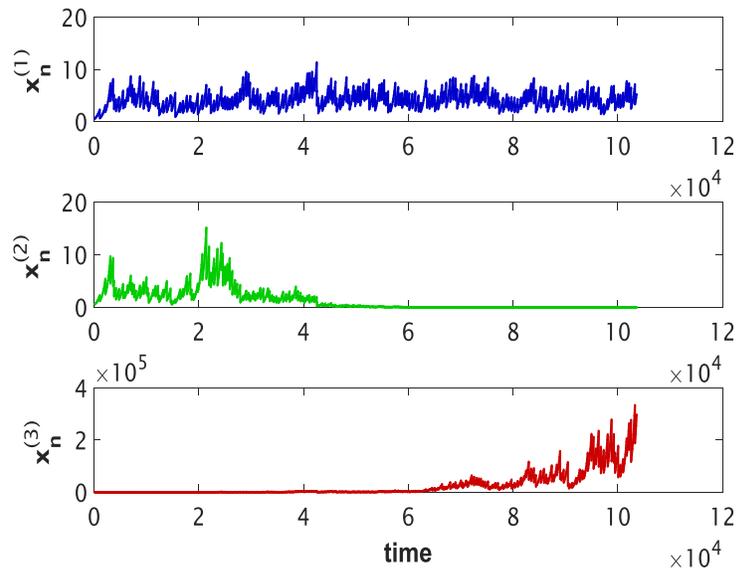

**Fig. SI-7-2. Instability of multi-component dynamics.** The traces of three components out of 50 in a simulation of our model are presented. These traces were generated under the assumption that $x^{(1)}$ controls cell division, and that the other components enslaved to the first, namely, components are assumed to grow with the same exponential rates, up to noise. It is seen that components that do not control cell division are not stable and can either decay to zero, as in the case of $x^{(2)}$, or diverge, as in the case of $x^{(3)}$.

### SI-7-C: Different strategies of cell-division control

For a multi-dimensional system of phenotype components interacting within the cell-cycle, one could hypothesize different control strategies. We considered models with noisy thresholds as division triggers, in analogy with the correlation inferred from the data for a single component. For one controlling component, the results are presented above and in the main text. For more than one controlling component, there are at least two possibilities: either the cell divides when at least one of the conditions is satisfied, or it divides when both are satisfied. We find that in both cases the entire system is stable to the noisy division events, and that the dynamical and statistical properties are maintained. A more in-depth understanding of the significance of multi-component division control is still lacking, and should be further investigated.

## SI-8: Peak-to-peak distance in ACF: theoretical prediction

Starting from our mapping model, Eq. (3) in the main text,

$$\ln\frac{x_{n+1}}{x^*} = (1-\beta)\ln\frac{x_n}{x^*} + \ln f_n + \phi^* + \xi_n,$$

we define $u_n = \ln(\frac{x_n}{x^*})$. Assuming that $\overline{\ln f_n} + \phi^* = 0$, we have the simple discrete Langevin equation

$$u_{n+1} = (1-\beta)u_n + \xi_n,$$

where $\xi_n$ is a noise term with zero mean and given variance.

We are interested in the probability of the event in which the current cell size forms a peak, i.e. it is larger than both the next and the previous values: $u_{n-1} < u_n > u_{n+1}$. Under the assumption of a Gaussian noise $\xi_n$, one may compute this probability as

$$P = \frac{1}{4\sqrt{\pi}}\int d\zeta\, e^{-\zeta^2}\, erfc^2\left(\sqrt{\frac{\beta}{2+\beta}}\,\zeta\right). \qquad \text{(Eq. SI-8-1)}$$

The derivation is worked out in detail in (7). The period of an oscillatory pattern is then approximated as the inverse of this probability. This is the expression for the mean time-interval between consecutive peaks $M(\beta)$, plotted in Fig. 6 of the main text.

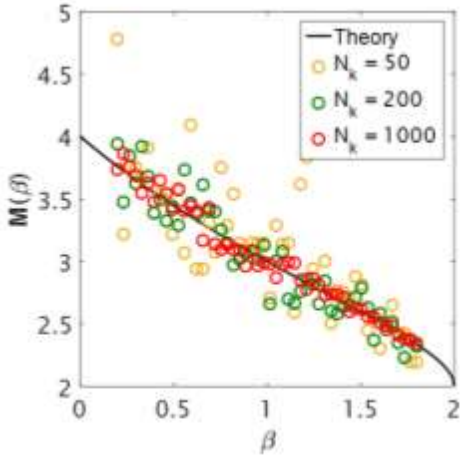

**Figure SI-8. Mean peak-to-peak distance $M(\beta)$ in ACFs: model simulations vs. theory.** The results of mapping simulations are compared to the theoretical prediction of Eq. (Eq. SI-8-1) (black line). Extending the length of the simulated traces, $N_k$, improves agreement with the theory (legend).

## SI-9: Experimental setup and stability

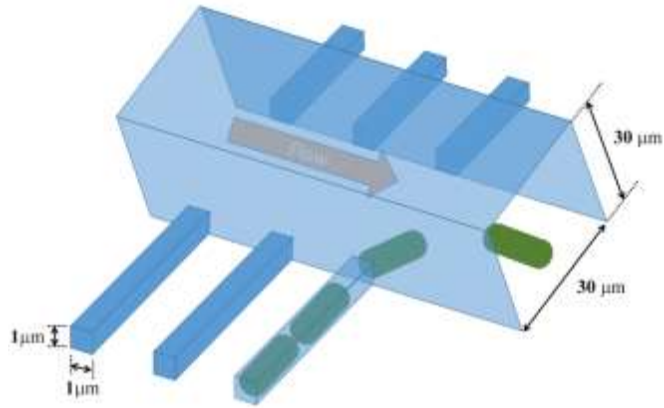

**Figure SI-9-1: Experimental Setup.** Bacteria (depicted in green) were trapped in an array of long micro-channels (1μm width 1 μm height), micro-fabricated in PDMS. The micro-channels were closed at one end and open at the other to large perpendicular channels (30 μm width 30 μm height), through which medium could be pumped in order to feed the trapped bacteria and to allow growth for many generations along the micro-channels.

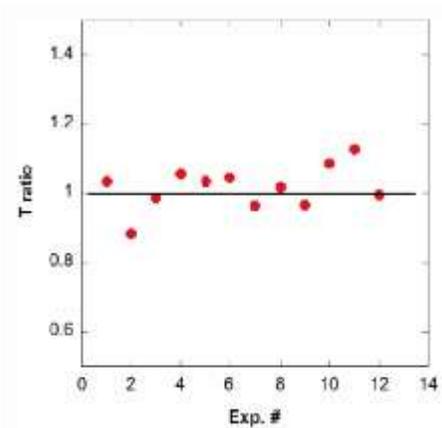

**Figure SI-9-2. Stability validation of the experimental procedure.** Shown are the average cell-cycle times for several traces during the first half of the trace, divided by the same average over the second half of the trace. The ratio for different traces is presented in the figure. We have chosen only traces that are longer than 100 generations for this test. The lack of trend indicates that the system was very stable over time, and that the experimental procedure did not influence our measurements.

## SI-10: Analysis of data from another lab

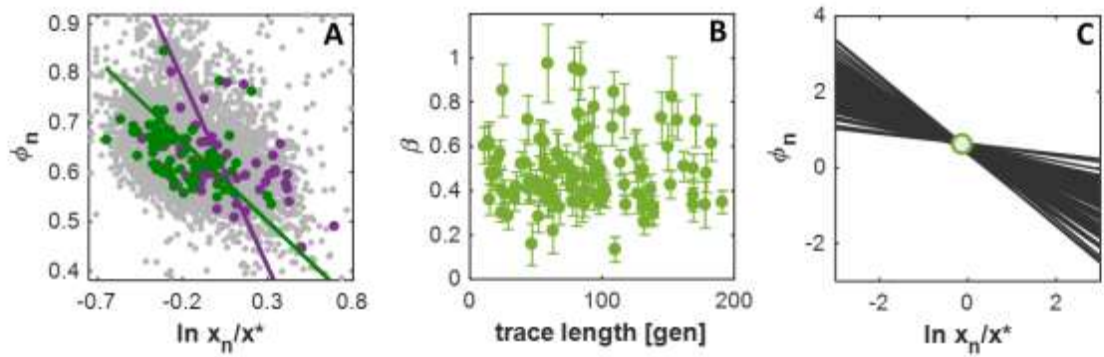

**Fig. SI-10: Analysis of individual traces from online available data.** (A) Pooled data from many traces (grey), and two individual traces highlighted in color. Plotted are the exponential accumulation per cycle as a function of log cell size at the cycle start. (B) Value of best fit slopes $\beta$ as a function of trace length in generations. (C) Best linear fits for a collection of traces exhibits an invariant pivot point. This figure was compiled from all mother-cell traces with >40 and <110 cycles in the directory 'E. coli MG1655 (CGSC 6300)\20090512\' from the dataset accompanying Ref. (8) (a total of 107 individual traces).